\documentclass{article}
\usepackage{graphicx}
\title{Simple approach to the mesoscopic open electron resonator: \\ Quantum
current oscillations }
\author{Constantino A. Utreras D\'{\i}az (cutreras@uach.cl) \\ Instituto de F\'{\i}sica,
Facultad de Ciencias \\ Universidad Austral de Chile \\ Campus Isla Teja s/n,
Casilla 567, Valdivia, Chile. \and J. C. Flores (cflores@uta.cl)
\\ Departamento de F\'{\i}sica, Facultad de Ciencias \\ Universidad de
Tarapac\'a,\\ Casilla 7D, Arica, Chile. \and Alejandro P\'erez Ponce
(aperez@fis.puc.cl)  \\ Facultad de F\'{\i}sica, Pontificia Universidad
Cat\'olica de Chile, \\ Avda. Vicu\~na Mackenna 4860, Casilla 306, Santiago
22 \\ Santiago Chile}

\date{\today}
\begin{document}
\maketitle
\begin{abstract}

The open electron resonator, described by Duncan et.al~\cite{DUNCAN}, is a
mesoscopic device that has attracted considerable attention due to its
remarkable behaviour (conductance oscillations), which  has been explained by
detailed theories based on the behaviour of electrons at the top of the Fermi
sea. In this work, we study the resonator using the simple quantum quantum
electrical circuit approach, developed recently by Li and
Chen~\cite{LI-CHEN}. With this approach, and considering a very simple
capacitor-like model of the system, we are able to theoretically reproduce
the observed conductance oscillations. A very remarkable feature of the
simple theory developed here is the fact that the predictions depend mostly
on very general facts, namely, the discrete nature of electric charge and
quantum mechanics; other detailed features of the systems described enter as
parameters of the system, such as capacities and inductances.

\end{abstract}

\maketitle
%
% Definiones utiles
%
\newcommand{\Df}{\phi_0}
\newcommand{\Dc}{\Delta_C}
\newcommand{\Ds}{\Delta_S}
\newcommand{\Dcp}{\Dc^{\prime}}
\newcommand{\Dsp}{\Ds^{\prime}}
\newcommand{\Cp}{C^{\prime}}
\newcommand{\Sp}{S^{\prime}}
\newcommand{\CM}{C_{(-)}}
\newcommand{\CP}{C_{(+)}}
\newcommand{\SM}{S_{(-)}}
\newcommand{\SP}{S_{(+)}}
\newcommand{\CMp}{C^{\prime}_{(-)}}
\newcommand{\CPp}{C^{\prime}_{(+)}}
\newcommand{\SMp}{S^{\prime}_{(-)}}
\newcommand{\SPp}{S^{\prime}_{(+)}}
\section{Introduction}
\begin{figure}
\includegraphics[width=5.0 cm]{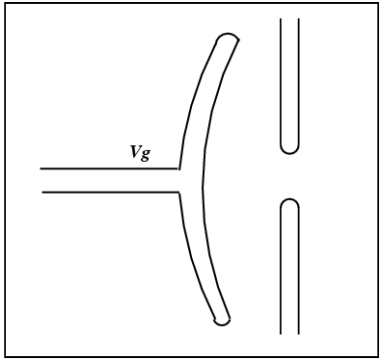}%
\caption{\label{FigRes}A drawing of the open resonator of Duncan et.
al.~\cite{DUNCAN}, showing the reflector (left) and the two point
contacts. }
\end{figure}

The field of mesoscopic physics deals with the frontier between
classical and quantum physics. However, this frontier is not
sharply defined, and every case needs special attention. Phenomena
like persistent currents~\cite{IMRY}, Coulomb blockade,
magneto-resistance fluctuacions and current
magnification~\cite{BENJAMIN1,BENJAMIN2,FLORES,FLORES-UTRERAS} are
usually studied in this area. The open electron resonator is a
device that has attracted some interest
recently~\cite{KATINE,SPECTOR1,SPECTOR2}. It consists of an open
cavity in a two-dimensional electron gas, into which electrons can
be injected via a quantum point contact, as described by Duncan
et. al~\cite{DUNCAN}, and shown in Figure~\ref{FigRes}. Some very
interesting properties have been found for this device, such as
the conductance peaks observed upon changing the cavity length by
$\lambda_f /2$ (half the 2D Fermi wavelength). Recently, Duncan
et. al~\cite{DUNCAN} have presented a series of conductance
measurements, showing that the measured conductance through the
point contact possesses a series of maxima, as a function of the
gate voltage $V_G$. These measurements have been performed both at
zero magnetic field ($B$), and as a function of it.

In this work, we present an alternative approach that allows us to
explain the observed conductance peaks, as a function of the gate
voltage $V_G$, at zero magnetic field, $B=0$. This approach is
related to the analogy between the mesoscopic device and a quantum
electrical circuit with charge discreteness, and it will be
discussed in the next section. In section \ref{seccion3} we study
the properties of the energy spectrum of our model.

\section{Our model}
\label{seccion2} We  start by observing that the open resonator of
reference~\cite{DUNCAN} is similar to a three-conductor system, in
which one of them, the gate, is kept at voltage $V_G$, while
current flows between the other conductors. We complete our model
device by assigning to it a self-inductance $L$. Classically, the
circuit may be described as a sort charged $L-C$ circuit. We
assume then that a current $I$ flows between the two charged
conductors (with charges $Q$ and $-Q$); further, for simplicity,
assume a plane geometry, and compute the classical electrostatic
energy, obtaining

\begin{equation}
U = C_1 V_g^2 + \frac{Q^2}{2 C_2} + \alpha Q V_g ,
\end{equation}
where $C_1$ and $C_2$ are capacities of the system, and $\alpha$ is a
geometry-related number of order unity. Note that this fact will enable us
to obtain some rough numeric estimates, based on the physical dimensions of the
device, which are in reasonable agreement with observations, as it will be
shown later.

As it has been proposed by Li and Chen~\cite{LI-CHEN}, and later by
Flores~\cite{FLORES}, and Flores and Utreras~\cite{FLORES-UTRERAS}, we treat
our mesoscopic system as a quantum electrical circuit, with quantized charge.
The classical Hamiltonian $H_{cl}$ of the (circuit) model system, written in
terms of the canonically conjugate variables $Q$ and $\Phi$ is

\begin{equation}
H_{cl} = \frac{\Phi^2}{2 L } + U = \frac{\Phi^2}{2 L } +  C_1 V_g^2 +
\frac{Q^2}{2C_2} + \alpha Q V_g .
\end{equation}

To quantize the system, the variables $Q$ and $\Phi$ are replaced by the
operators $\widehat Q$ and  $\widehat \Phi$. Further, to recover the
quantization of charge within this electrical circuit approach, we introduce
the replacement~\cite{LI-CHEN,FLORES,FLORES-UTRERAS}

\begin{equation}
\label{Eq-cambio}
\widehat{\Phi }\rightarrow \frac{2\hbar }{q_{e}}\sin \left(
\frac{q_{e}}{2\hbar}  \widehat{\Phi }\right),
\end{equation}
where $q_e$ is the quantum of charge. We remark that, if charge
discreteness is neglected, the operator $\widehat \Phi$ may be
directly identified with the magnetic flux operator, and therefore
directly related to the current; however, when one introduces
charge discreteness via the replacement above, the simple relation
to the current is lost, therefore, after
replacement~(\ref{Eq-cambio}) the flux operator becomes the
pseudo-flux. This pseudo-flux operator satisfies the usual
commutation relation$[\widehat Q, \widehat \Phi] = i \hbar$.
Notice that, given the complexity of dealing with an operator such
as the one above (\ref{Eq-cambio}), it is simpler to work in the
so-called pseudo-flux representation, in which the operator
$\widehat \Phi$ is replaced by its eigenvalue $\phi$, while the
charge operator is given by $\widehat Q = i \hbar\partial /
\partial \phi$. In this way, the resulting hamiltonian is given by

\begin{equation}
\label{Hamiltonian}
\widehat H =  \frac{2 \hbar^2}{q_e^2 L} \sin^2 ( q_e \widehat \phi /2 \hbar) +
C_1V_g^2 + \frac{\widehat Q^2}{2 C_2} + \alpha \widehat Q V_g.
\end{equation}

The hamiltonian operator above constitutes our starting point, and  our working
hypothesis. The parameters of our theory, particularly $L$ and $C$, are related
to the geometry of the system, but are hard to compute for a given experimental
system, therefore, they should be deduced from experimental observations.

\section{Energy spectrum}
\label{seccion3}

To study the properties of our system, define first $\beta = (C_1
- \alpha^2 C /2 ) $, $C=C_2$, and $a = \alpha C V_g/\hbar$, and
rewrite our hamiltonian as

\begin{equation}
\widehat H = \beta V_g^2  + \frac{2 \hbar^2}{q_e^2 L} \sin^2 ( q_e
\phi /2 \hbar) + \frac{\hbar^2}{2C} ( i \frac{d}{d\phi} + a)^2 .
\end{equation}

It is observed that, in this representation, the magnetic energy
term acts like a periodic potential in the variable $\phi$, while
the electrostatic energy term here becomes a kind of 'kinetic'
term, i.e., formally, the terms reverse their meaning, when one
compares with a 'regular' Schr\"odinger equation. The term $\beta
V_g^2$ is just a constant, and it will be eliminated from the
hamiltonian. To make our analogies more clear, define a periodic
'potential energy function' $V(\phi)$, with period $\phi_0 = 2 \pi
\hbar /q_e = h /q_e$

\begin{equation}
V(\phi) = \frac{2 \hbar^2 }{q_e^2 L } \sin^2 (q_e \phi /2 \hbar ) =
\frac{ \hbar^2 }{q_e^2 L } \left\{ 1 - \cos(q_e \phi / \hbar) \right\}.
\end{equation}

The Schr\"odinger equation $H \Psi = E \Psi$ may now be written as

\begin{equation}
\label{EqModel}
 \left\{ V (\phi ) + \frac{\hbar^2}{2 C} ( i \frac{d}{ d\phi} + a )^2
\right\} \Psi (\phi) = E \Psi (\phi).
 \end{equation}
The wavefunction $\Psi (\phi)$ satisfies the strong periodicity condition

\begin{equation}
\label{EqPeriodicity} \Psi ( \phi + \phi_0 ) = \Psi(\phi),
\end{equation}
this condition results from the fact that quantization in the charge space
implies periodicity in the (pseudo) flux space, as shown by Li and
Chen~\cite{LI-CHEN}.

Observe that the hamiltonian $\widehat H$ above commutes with the
discrete translation operator $\widehat T_{m \phi_0} \Psi(\phi) =
\Psi (\phi + m \phi_0 )$, for integer $m$; therefore, for any
value of $a$, Bloch's theorem holds, specifically, the energy
$E(a)$ should be a periodic function of $a$, labeled by a 'band
index' $n$, further, the parameter $a$ should play the role of the
wavenumber in Bloch theory; the period for $a$ is $\Delta a =
q_e/\hbar$, and therefore, the energies are periodic functions of
the gate voltage $V_G$, with period $\Delta V_g = q_e/ \alpha C $
(recall that $\alpha$ is a number that we take as unity from here
on).

To proceed with the numerical calculations, we should get some
estimates of the parameters of the theory, namely, the values of
the capacity $C$ and the inductance $L$. These may be obtained
from the knowledge of the linear dimensions of the system, i.e.,
of the order of $1 \mu m$, and using the classical formula $ C =
\epsilon_0 l b/d$, in which $l$ is taken as the lateral dimension
of the cavity, $d$ the distance of the contact points and $b$ is
the thickness of the 2D electron gas in which the device is
immersed, now since $l \approx d$, and $b$ is a fraction of $a \mu
m$, we obtain $\Delta V_G \approx 0.02 (Volts)$, a reasonable
estimate of the effect. As to the estimate for $L$, it is a
parameter related to the size of the circuit that contains the
device, i.e., a relatively large number $L \approx \mu_0 D$. As a
measure of this, we have taken the ratio of the magnetic potential
energy, $\phi_0^2/L$ to the electric energy $q_e^2/2C$ to be of
order unity, for the numerical calculations described here.

Notice that we have estimated the capacity of
the system using the classical formula, but to include small-size
corrections it may be more appropriate to obtain tha capacity from a
microscopic theory, or, at least from Thomas-Fermi theory.

\begin{figure}
\includegraphics[width=10.0 cm]{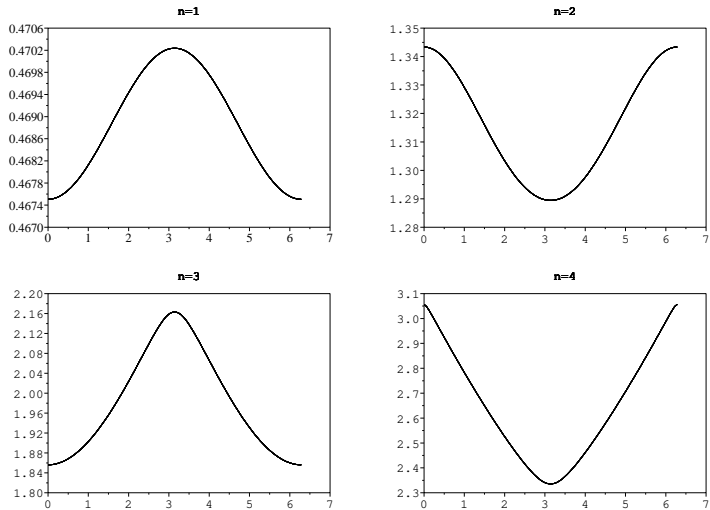}%
\caption{\label{Ebands} Computed energy eigenvalues (first bands) as a function
of the gate voltage $V_G$. The ordinate is the energy, on the dimensionless 
energy scale $(\phi_0^2/2L)/(q_e^2/2C)$, the abcisa is the dimensionless parameter
$a \phi_0 = (2 \pi C/q_e) V_G$. Observe that the ground state 'band' has a very small
oscillation amplitude, if shown on the same scale than the other curves, it
would appear completely flat. }
\end{figure}

In Appendix~\ref{Apendice}, we discuss the procedure used to find
the energy eigenfunctions $u_{E}(\phi)$ and their corresponding
energies for our hamiltonian. The solutions are expressed in terms
of two normalized and mutually orthogonal functions $C_E(\phi)$ y
$S_{E}(\phi)$. The eigenvalue conditions is expressed by
eq.(~\ref{EqEigen}), which shows explicitly the periodicity of the
energies as a function of $V_G$.

We have computed the eigenfunctions $u = u_{E}(\phi) = u_{n,V_G}
(\phi)$, by numerically solving the eigenvalue
equation (Eq.~\ref{EqEigen}, in Appendix), for the lowest
eigenstates, as a function of the parameter $a = C V_g/\hbar$. 
The full wavefunction $u(\phi)$ is given by the linear combination

\begin{equation}
u_E (\phi) = A  C_E(\phi) + B S_E(\phi),
\end{equation}
with coefficients given by

\begin{eqnarray}
A & = & \frac{\Delta_S}{\sqrt{|\Delta_C|^2 + |\Delta_S|^2}}
\nonumber \\ B & = & - \frac{\Delta_C}{\sqrt{|\Delta_C|^2 +
|\Delta_S|^2}} . \nonumber
\end{eqnarray}

The results of our energy calculations are displayed in Figure~\ref{Ebands},
which shows that the energies are distributed in
energy bands with the periodic structure already described (period
$\Delta a =2\pi/\Df$). Observe that the first band has very small width,
which depends on the relation of electric to magnetic energy.

\section{Conductance Oscillations}

From a general point of view, we may expect the same periodicity in the
conductance as that of the energy.  If we assume  $A$ as the probability per
unit of time of decaying transitions between two levels $(n,n')$ then the
emission power is $A( E^0 (n)-E^0 (n'))$, which must be proportional to the
conductance $G$ (the proportionality constant is the square of the external
bias voltage). Since the energy spectrum is periodic in $\Delta V_g = q_e/
\alpha C$, so is the conductance.

We have shown that the general features of the energy sprectrum do not depend on the
value of the inductance $L$, in the sense that, as long as $L \ne 0$, the energy
spectrum will display the $V_G$ periodicity. However, for small values of $L$, this
behaviour will not be seen, since the amplitude of the energy oscillations will be
very small. This is actually observed (see figure~\ref{Ebands}) for the lowest energy 'band', even
for moderate values of $L$, in other words, the details of the energy
spectrum do depend on $L$, but its general properties, such as the periodicity
depend only on the capacity of the system, as shown by the experiments.

We remark that our approach is able to explain the experimentally
observed~\cite{DUNCAN} oscillatory behaviour of $G$, as a function
of $V_G$, but the linear dependence of the current on $V_G$, is
not directly explained by our approach, or, to put it
differently, the present theory does not describe the linear
dependence of the current on the gate voltage. We observe that, 
for fixed $V_G$, we may change the capacity of
the system, $C^{\prime} = \alpha C_2$, by changing one of the
linear dimensions of the cavity($l$, say), while keeping constant
both the distance from the contact points constant ($d$), and the
other linear dimension ($b$); it is clearly seen then that the
current is also periodic in $l$, i.e., since we may write for
$C^{\prime}$

\begin{equation}
C^{\prime} = \epsilon_0 \alpha  l b / d ,
\end{equation}
in this way, the oscillatory behaviour of the current (and, hence, of the
conductance too) as a function of the resonator cavity length may also be
explained.

Now we make an estimate of the size of the effect, as suggested by
our theory. Note that in $C^{\prime}$ above, the area $l \times b
$ is the product of the linear dimensions of the reflector, while
$d$ is the distance to the contact points. We see that two of
these distances are of order $ 1 \mu m$, therefore, $C^{\prime}
\approx \epsilon_0 \times b$ ($b$ being the perpendicular
dimension to the 2D electron gas, of order $1 \mu m $, then
$\Delta V_g \approx 0.02 (V)$, a pretty good answer, given the
rough estimate for the capacity $C^{\prime}$ used here.

\section{Discussion}

This article offers an alternative point of view on the open electron
resonator, while recognizing that there exists an accepted explanation of the
effect. Nevertheless, our alternative description is very simple,
powerful and it does not require elaborate assumptions. It is a new description
in terms of elements like capacitors and inductances, which is not too widely
different  from more usual treatments for mesoscopic systems~\cite{DATTA}.

It is also worth considering the issue of the spatial dimensionality of the
theory. A casual glance of our equations gives one the impression that the
theory is one-dimensional. This is not the case, since the theory is expressed
in terms of charge-flux variables, and no reference is made to the
dimensionality of the devices it describes; in fact, such spatial
dimensionality only appears through the parameters of the theory, such as
capacities and inductances, just as it happens in classical electrical
circuits. It is worth noticing that such theory is more correctly described as
''topological'', and not ''geometrical. Therefore, a correct use of the theory
may help in sheding some light in other mesoscopic systems in study today.

\section{Acknowledgements}

J. C. Flores and C. A.Utre\-ras D\'{\i}az acknowledge support from FONDECYT
(Grant No. 1040311), C. A.Utre\-ras D\'{\i}az acknowledges support from
Universidad Austral de Chile (DIDUACh Grant No. S-2004-03), and A.
P\'erez Ponce acknowledges the kind hospitality provided by the
Instituto de F\'{\i}sica, Universidad Austral de Chile, and
partial support from DIDUACh.

\appendix
\section{Solution of the Eigenvalue equation} \label{Apendice}

Consider now the (numerical) solution of the resonator
equation~(\ref{EqModel}), for the eigen-function $\Psi(\phi)$ and
eigen-energies, as a function of the parameter $V_G$ ($a$), with
the periodic boundary conditions~(\ref{EqPeriodicity}). As
indicated previously, the wavefunction $\Psi(\phi)$ may be written
as

$$ \Psi(\phi) = e^{i a \phi} u(\phi)$$

This replacement has the property that it eliminates the constant
$'a'$ from the Schr\"odinger equation for the unknown function $
u(\phi)$. This wavefunction is not periodic, but it satisfies a
simple Bloch-type boundary condition (which follows from the
periodicity of $\Psi$).

\begin{equation}
u(\phi + \phi_0) = e^{-ia\phi_0} u(\phi)
\end{equation}

Now, to actually find numerical solutions of the Schr\"odinger
equation, and their corresponding energies, we proceed as follows.
For each value of the energy $E$, we find two independent
numerical solutions, which we call $C(\phi)$ and $S(\phi)$. For
example, for a fundamental period with endpoints at $\phi = 0$ and
$\phi = \phi_0$, these functions may be chosen to satisfy
\begin{eqnarray}
C_E(\phi_0/2) = 1 & \Cp_E (\phi_0/2) = 0 \\ S_E(\phi_0/2) = 0 &
\Sp_E(\phi_0/2) = 1
\end{eqnarray}

The wavefunction $u$ may now be expressed as the linear
combination

\begin{equation}
u_E (\phi) = A \cdot C_E(\phi) + B\cdot S_E(\phi).
\end{equation}

The boundary conditions are now

\begin{eqnarray}
u_E(\phi_0 ) = e^{-i a \phi_0} u_E(0) \nonumber\\
 u_E^{\prime}(\phi_0) = e^{-i a \phi_0} u_E^{\prime}(0),
\end{eqnarray}
which lead to two linear equations for the coefficients $A$ and
$B$,

\begin{eqnarray}
\Dc A + \Ds B & =  &  0 \nonumber \\ \Dcp A + \Dsp B & = & 0 ,
\end{eqnarray}
in which the coefficients are

\begin{eqnarray}
\Dc & = & e^{ia \phi_0/2} C(\phi_0) - e^{-i a\phi_0/2} C(0 )
\nonumber \\ \Ds & = & e^{i a \phi_0/2} S(\phi_0) - e^{-i a
\phi_0/2} S(0) \nonumber \\ \Dcp & = & e^{i a \phi_0/2}
C^{\prime}(\phi_0) - e^{-i a \phi_0/2} C^{\prime}(0) \nonumber
\\ \Dsp & = & e^{i a \phi_0/2} S^{\prime}(\phi_0) -
e^{-i a \phi_0/2} S^{\prime}(0)
\end{eqnarray}

The eigenvalue equation for the energies is given by the vanishing
of the determinant of the coefficients,
\begin{equation}
G (E) = \Dc \Dsp - \Ds \Dcp = 0
\end{equation}

When, as it happens in our case, the potential is symmetric with
respect to the midpoint of the interval, the functions $C$ and $S$
have definite parity about that point, i.e. $C(\phi_0) = C(0)$ and
$S(\phi_0 ) = -S(0)$, the derivatives have the opposing parity,
$C^{\prime}(\phi_0) = -C^{\prime}(0)$ and $S^{\prime}(\phi_0) =
S^{\prime}(0)$, therefore, defining $\theta = a \Df /2$, we obtain
a simplified eigenvalue equation,

\begin{equation}
G(E) = \sin^2(a\phi_0/2) C(0) S^{\prime}(0) + \cos^2(a \phi_0/2)
S(0) C^{\prime}(0) = 0 \end{equation}

This equation is convenient for numerical work, since the
functions $C$ and $S$ are easily computed numerically; to find the
eigenvalues $E$, we fix $a$ and compute $G(E)$, the zeros
correspond to the eigenstates.

For the general case (non symmetric), it is useful to define the
coefficients

\begin{eqnarray}
C_{(-)} & = & C(\Df) - C(0)\nonumber \\ C_{(+)} & = & C(\Df) +
C(0)\nonumber
\\ S_{(-)} & = & S(\Df) - S(0) \nonumber \\ S_{(+)} & = &  S(\Df) + S(0) \nonumber  \\ \Cp_{(-)} &
= &  \Cp (\Df) - \Cp(0) \nonumber \\ \Cp_{(+)} & = &  \Cp(\Df) +
\Cp(0) \nonumber \\ \Sp_{(-)} & = & \Sp(\Df) - \Sp(0) \nonumber
\\ \Sp_{(+)} & = & \Sp(\Df) + \Sp(0)\nonumber  .
\end{eqnarray}

With this, the coefficients of the eigenvalue equation become

\begin{eqnarray}
\Dc & = & \CM \cos(\theta) + i \CP \sin(\theta) \nonumber \\ \Ds &
= & \SM \cos(\theta) + i \SP \sin(\theta) \nonumber \\ \Dcp & = &
\CMp \cos(\theta) + i \CPp \sin(\theta) \nonumber
\\ \Dsp & = & \SMp \cos(\theta) + i \SPp \cos(\theta)
\end{eqnarray}

The eigenvalue equation is purely real (its imaginary part  $\Im [
G(E) ] = 0$ vanishes identically since the Wronskian -$W[C,S]$- of
two solutions is a constant):

\begin{equation}
\label{EqEigen}
 G (E) = \left( C_{(-)} \Sp_{(-)} - S_{(-)} \Cp_{(-)}
\right) \cos^2(\theta) + \left( S_{(+)} \Cp_{(+)} - C_{(+)}
\Sp_{(+)} \right) \sin^2(\theta).
\end{equation}

\end{document}